# On Virk's Asymptote

Trinh, Khanh Tuoc

Institute of Food Nutrition and Human Health

Massey University, New Zealand

*K.T.Trinh@massey.ac.nz*

## Abstract

Virk's asymptote is shown to be similar in nature to Karman's buffer layer profile and does not represent a new log-law with a modified mixing-length. It is simply part of the wall layer velocity profile but is extended because of the increase in wall layer thickness in drag reduction flows. The friction factors at the maximum drag reduction asymptote correspond to velocity profiles consisting of a wall layer and a law of the wake sub-region. Maximum drag reduction results in a the suppression of the law of the wake and full relaminarisation of the flow.

## Introduction

The phenomenon of drag reduction DR was discovered accidentally by Toms (1949) who found that the addition of a small amount of polymers to water reduced the turbulent friction factor substantially. The most notable application is in the Alaska Pipeline System where it helps decrease power requirements and increase design throughput (Burger, Chorn, & Perkins, 1980) and friction drag on vessel hulls (National Research Council, 1997). Kawaguchi, Li, Yu, & Wei (2007) report that 70% of the pumping power used to drive hot water in primary pipelines or district heating systems was saved by adding only a few hundred ppm of surfactant into the circulating water. The significant potential of this technology in the reduction of energy usage in the face of dwindling resources as well as reduction in environmental pollution has encouraged a large number of studies on viscoelastic flows. These studies are also made to add further insight into the mechanics of turbulence and polymer behaviour. Excellent reviews have been published in the last 40 years (Gyr &

Bewersdorff, 1995; Hoyt, 1972; Landahl, 1973; Liaw, Zakin, & Patterson, 1971; J. Lumley, 1969; J. L. Lumley, 1973; McComb, 1991; Nieuwstadt & Den Toonder, 2001; Virk, 1975; White & Mungal, 2008).

Polymer addition is not the only way to achieve DR. It has been known for some time that the drag force of a Newtonian fluid in turbulent flow past a surface can be reduced significantly by the addition of small riblets on the wall in the streamwise direction e.g. (P.R. Bandyopadhyay, 1986; Bechert & Hage, 2006). The principle is copied from observations of shark skin and now widely applied, for example in the design of swimsuits to improve swimmers performance(Orlando news, 2004).

Another technique which is copied from the swimming techniques of dolphins is to use compliant surfaces (P.R. Bandyopadhyay, 1986; P. R. Bandyopadhyay, 2005; Choi et al., 1997). The resultant wall motion is a uniform wave travelling downstream (Fukagata, Kern, Chatelain, Koumoutsakos, & Kasagi, 2008). In wind tunnel measurements Lee, Fisher, & Schwarz (1995) showed that the flow-induced surface displacements resulted in a reduction in the growth rates of unstable Tollmien-Schlichting waves and a delay in the onset of turbulence.The generation or injection of bubbles near the surface is another exciting technique for drag reduction e.g. (McCormick & Bhattacharyya, 1973; Merkle & Deutsch, 1989; Mohanarangam, Cheung, Tu, & Chen, 2009; Sanders, Winkel, Dowling, Perlin, & Ceccio, 2006). The effect is attributed to the deformable properties of bubble surfaces e.g. (Oishi, Murai, Tasaka, & Yasushi, 2009). Recently Watanabe and Udagawa (2001) showed significant DR by coating the wall surface with a super water repellent paint SWR and (Fukuda et al., 2000) introducing air which forms a thin lubricating film.

These technologies are being actively studied applied in both air and water transports e.g. (P. R. Bandyopadhyay et al., 2005; Bushnell, 2003; Choi, et al., 1997; Davies, Carpenter, Ali, & Lockerby, 2006; McCormick & Bhattacharyya, 1973; Reneaux, 2004).

A number of physical phenomena have been observed in DR flows. However a majority of the studies are focused on the Toms phenomenon through the addition of polymers even though DR can also be achieved by the addition of cationic surfactants (Chara, Zakin, Severa, & Myska, 1993; Gasljevic, Aguilar, & Matthys, 2001; Myska & Zakin, 1997) who point out that the mechanisms of DR in solutions with polymer and surfactant show clear differences.

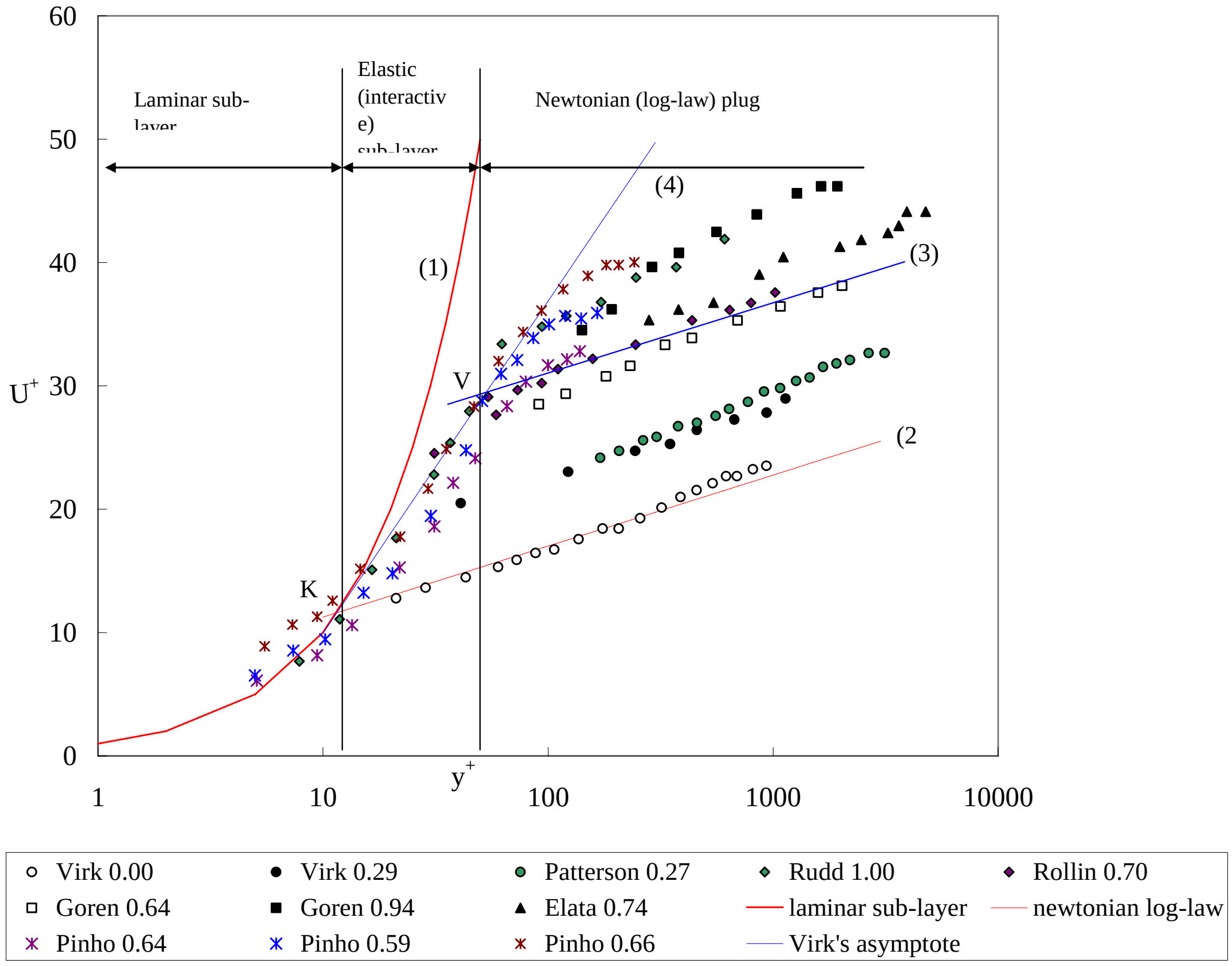

Figure 1. Velocity profiles of polymer solutions. Numbers in the legend refer to the extent of DR for the particular fluid flow shown.

All authors agree that there is a substantial thickening of the wall layer. Thus the mean velocity distribution is modified and the shear in the boundary layer is redistributed (White & Mungal, 2008). At low DR the Prandtl-Nikuradse log law is pushed further away from the wall and therefore the slope of the Karman buffer layer (the region between the Prandtl laminar sub-layer and the log law) is increased as shown in Figure 1 where the original data used by Virk (1975) has been replotted with additional newer data by Pinho and Whitelaw(1990). Sibilla and Beretta (2005) report from their DNS analysis that the mean vortex in the polymer flow is weaker less tilted and inclined, while its length and radius are higher. Cross-flow velocity fluctuations near the wall drop consistently and the Reynolds stresses in the buffer layer are reduced. Similar observations are found in experimental work e.g. (Dubief, Terrapon, Shaqfeh, Moin, & Lele, 2004; Hetsroni, Zakin, & Mosyak, 1997; Itoh, Imao, &

Sugiyama, 1997; Ptasinski, Nieuwstadt, van den Brule, & Hulsen, 2001; White, Somandepalli, & Mungal, 2004) and numerical simulations e.g. (Akhavan & Lee, 2007; Min, Cho, & Jung, 2003; Pereira & Pinho, 1994; Ptasinski et al., 2003; Stone, Roy, Larson, Waleffe, & Graham, 2004; Vaithianathan & Collins, 2003) and Dubief, Terrapon, Shaqfeh, Moin, & Lele (2004).

Polymer solutions are known to be viscoelastic and many authors naturally sought to explain DR in terms of the polymer relaxation time (Fang, Hasegawa, Sorimachi, Narumi, & Takemura, 1998). Darby and co-workers (Darby & Chang, 1984; Darby & Pivsaart, 1991) provided new scaling criteria by including a relaxation time and collapsed viscoelastic and non-elastic friction factors. Dimensional analysis introduced two parameters, the Deborah number (Seyer & Metzner, 1967) and the Weissenberg number. Both are dimensionless ratio of time scales: a typical polymer response time, usually the relaxation time which for linear polymers can be related to the size and number of monomers (Flory, 1971) and a representative time scale of the turbulent process. The difficulty lies in the proper choice of this time scale. Some authors use $T = v / u_*^2$ (White & Mungal, 2008).

More detailed analyses of the mechanics of turbulent viscoelastic flow involve considerations of the energy required to stretch the polymer. Most authors agree that this process occurs in the buffer layer and alters the flow dynamics. White and Mungal (op.cit.) identify two schools of thoughts. The first focuses on viscous effects and the second on elastic effects. The first school proposes that the effect polymer stretching is to increase the effective viscosity of the solution which in turn suppresses turbulent fluctuations. The second school argues that the onset of DR occurs when the elastic energy stored by the partially stretched polymer becomes comparable to the kinetic energy in the buffer layer at some scale larger than the Kolmogorov scale, the scale of the energy dissipating eddies. The forces linked with polymer stretching are extracted from laser velocimetry, PIV and DNS experiments by expressing the total local averaged stress as

$$\tau = \mu \frac{\partial U}{dy} - \rho \overline{u'v'} + \tau_{xy}^{p} \qquad (1)$$

White and Mungal state that “the increasing role of the polymer shear stress $\tau_{\kappa y}^{p}$ is perhaps the most interesting and controversial characteristic of HDR flows".

While a full consensus on the mechanisms of DR is still not available, the existence of a maximum level of achievable DR called Virk's asymptote (Virk, Mickley, & Smith, 1970) is widely accepted. Virk et al divide the flow field into three regions:

1. A laminar sub-layer first proposed by Prandtl (1935) where the velocity follows the relation

   $$U^+ = y^+ \tag{2}$$

The velocity and distance have been normalised with the wall parameters $U^+ = U/u_*$, $y^+ = yu_*/\nu$, $u_* = \sqrt{\tau_w/\rho}$, $\tau_w$ is the average wall shear stress, $\rho$ the fluid density and $\nu$ the apparent kinematic viscosity.

2. A "Newtonian plug" (line 2) where the velocity profile follows the Prantl-Millikan log-law . Despite Virk et al's nomenclature, the log-law also applies to non-elastic non-Newtonian fluids (K. T. Trinh, 2009)

   $$U^+ = 2.5\ln y^+ + 5.5 \tag{3}$$

3. An elastic or interactive sublayer which starts at the point K and results in an effective slip because it pushes the log-law plug away from the wall. It follows the equation

   $$U^+ = 11.7\ln y^+ - 17.0 \tag{4}$$

For Newtonian fluids, Virk et al argue that there are only two regions: the laminar sublayer and the log-law that intercept as the point K. An example of the shifted log-law is shown as line (3) in Figure 1 for the data of Goren and Norbury (1967)at a DR of 64%. In this case the log-law and line (4) intersect at the point V which is farther from the wall than point K. Virk (1975) proposed the intercept occurs when the time-averaged wall shear stress reaches a value characteristic of the polymer-solvent pair. The elastic sub-layer is only present in DR flows. Virk et al observed that the extend of the log-law decreases with increasing DR and simultaneously the elastic sub-layer increases. In their view, this asymptotic velocity profile is different from purely viscous (laminar) flow and they illustrate the difference by plotting the friction factor at maximum DR (Figure 2). Thus in their view, the asymptotic velocity profile represents a new log-law which is different from the Prandtl-Millikan log-law because it represents the interaction between the polymers and the viscous flow. This view has been adopted by many subsequent authors e.g.(McComb, 1991; McComb & Rabie,

1979).

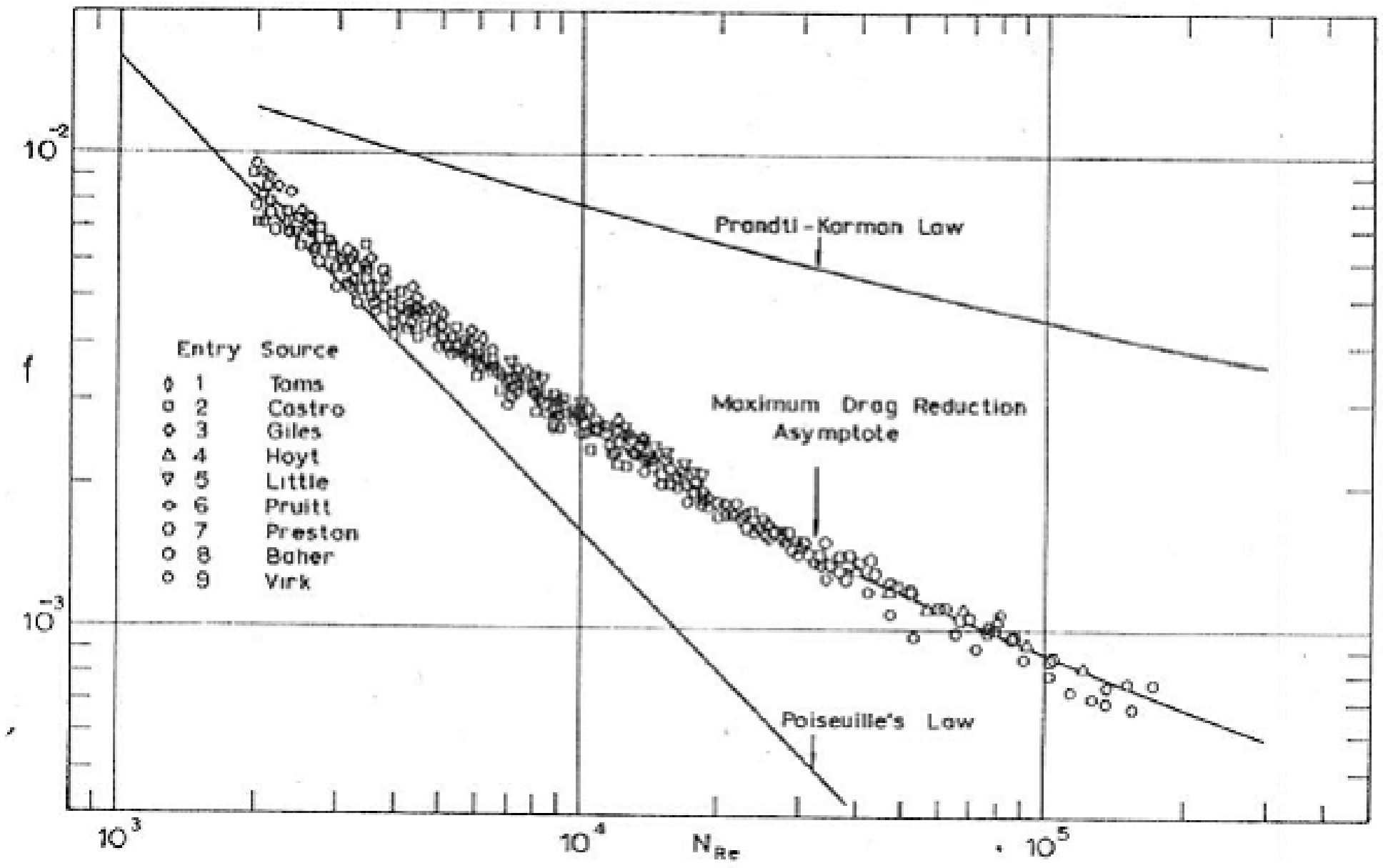

Figure 2 Friction factor at maximum drag reduction MDR (Virk, et al., 1970)

A number of authors have attempted to derive the Virk asymptote from theoretical considerations. Benzi et al.(2005) derived the functional form and the parameters that determine the maximum drag reduction asymptote from considerations of the balance of energy and momentum. Larson (2003) found the thicknesses of the laminar, buffer, and turbulent boundary layers as functions of dimensionless position Re, along the plate, as well as the drag on the plate for drag-reducing polymer as functions of the critical shear stress. At high Reynolds number (Re-x), around 109, a factor of six reduction in drag is the maximum reduction possible, corresponding to the maximum drag reduction (MDR) asymptote for pipe flow. Sher and Hetsroni (2008) proposed a new mechanistic model of a polymer molecule in a turbulent flow field. They argue that the dominant forces on a polymer fiber in the turbulent flow field are elastic and centrifugal. According to this model, an additional route of dissipation exists, in which eddy kinetic energy is converted to polymer elastic energy by the centrifugal elongation of the rotated polymer, which in turn is viscously damped by the surroundings, when the polymer relaxes. This mechanistic model can be accounted for as a turbulent scale alteration, instead of addition, which enables the classical dimensional analysis of a turbulent boundary layer to apply. Using this dimensional analysis with the equivalent altered scale correct-form velocity profiles are obtained, and Virk's asymptote and slope are predicted with no empirical constants.

A different interpretation of Virk’s asymptote is presented in this paper.

## Proposed visualisation

Turbulent flows are traditionally divided into an outer and an inner region e.g. Panton (1990). The inner region is further divided into a log-law sub-region and a wall layer where the effects of viscous diffusion of momentum predominate. In 1967, Kline et al. (1967) reported their now classic hydrogen bubble visualisation of events near the wall and ushered in a new area of turbulence research based on the so-called coherent structures. Despite the prevalence of viscous diffusion of momentum close to the wall, the flow was not laminar in the steady-state sense envisaged by Prandtl. Instead the region near the wall was the most active in the entire flow field. In plan view, Kline et al. observed a typical pattern of alternate low– and high-speed streaks. The low-speed streaks tended to lift, oscillate and eventually eject away from the wall in a violent burst. In side view, they recorded periodic inrushes of fast fluid from the outer region towards the wall. This fluid was then deflected into a vortical sweep along the wall. The low-speed streaks appeared to be made up of fluid underneath the travelling vortex. The bursts can be compared to jets of fluids that penetrate into the main flow, and get slowly deflected until they become eventually aligned with the direction of the main flow. This situation is found in studies of a vortex moving above a wall e.g. (Peridier, Smith, & Walker, 1991; Smith, Walker, Haidari, & Sobrun, 1991; Suponitsky, Cohen, & Bar-Yoseph, 2005; Walker, 1978) and results in the growth of the fluctuations and eventually eruption of the low-speed fluid beneath the vortex.

In other publications, the writer has shown that these events and structures are linked with specific solutions of the Navier-Stokes equations (K. T. Trinh, 2009). The instantaneous velocity in the sweep phase of the wall layer is divided into three components as shown in Figure 3.

If we draw a smooth line through this velocity trace so that there are no secondary peaks within the typical timescale of the flow $t_\nu$, we define a locus of smoothed velocity $\tilde{u}_i$ and fast fluctuations $u'_i$ of period $t_f$ relative this base line. The instantaneous velocity may be decomposed as:

$$u_i = \tilde{u}_i + u'_i \tag{5}$$

We may average the Navier-Stokes equations

$$\frac{\partial}{\partial t}(\rho u_i) = -\frac{\partial}{\partial x_i}(\rho u_i u_j) - \frac{\partial}{\partial x_i} p - \frac{\partial}{\partial x_i}\tau_{ij} + \rho g_i \tag{6}$$

over the period $t_f$ of the fast fluctuations. Bird, Stewart, & Lightfoot (Bird, Stewart, & Lightfoot, 1960) give the results as

$$\frac{\partial(\rho \tilde{u}_i)}{\partial t} = -\frac{\partial p}{\partial x_i} + \mu \frac{\partial^2 \tilde{u}_i}{\partial x_j^2} - \frac{\partial \tilde{u}_i \tilde{u}_j}{\partial x_j} - \frac{\partial \overline{u'_i u'_j}}{\partial x_j} \tag{7}$$

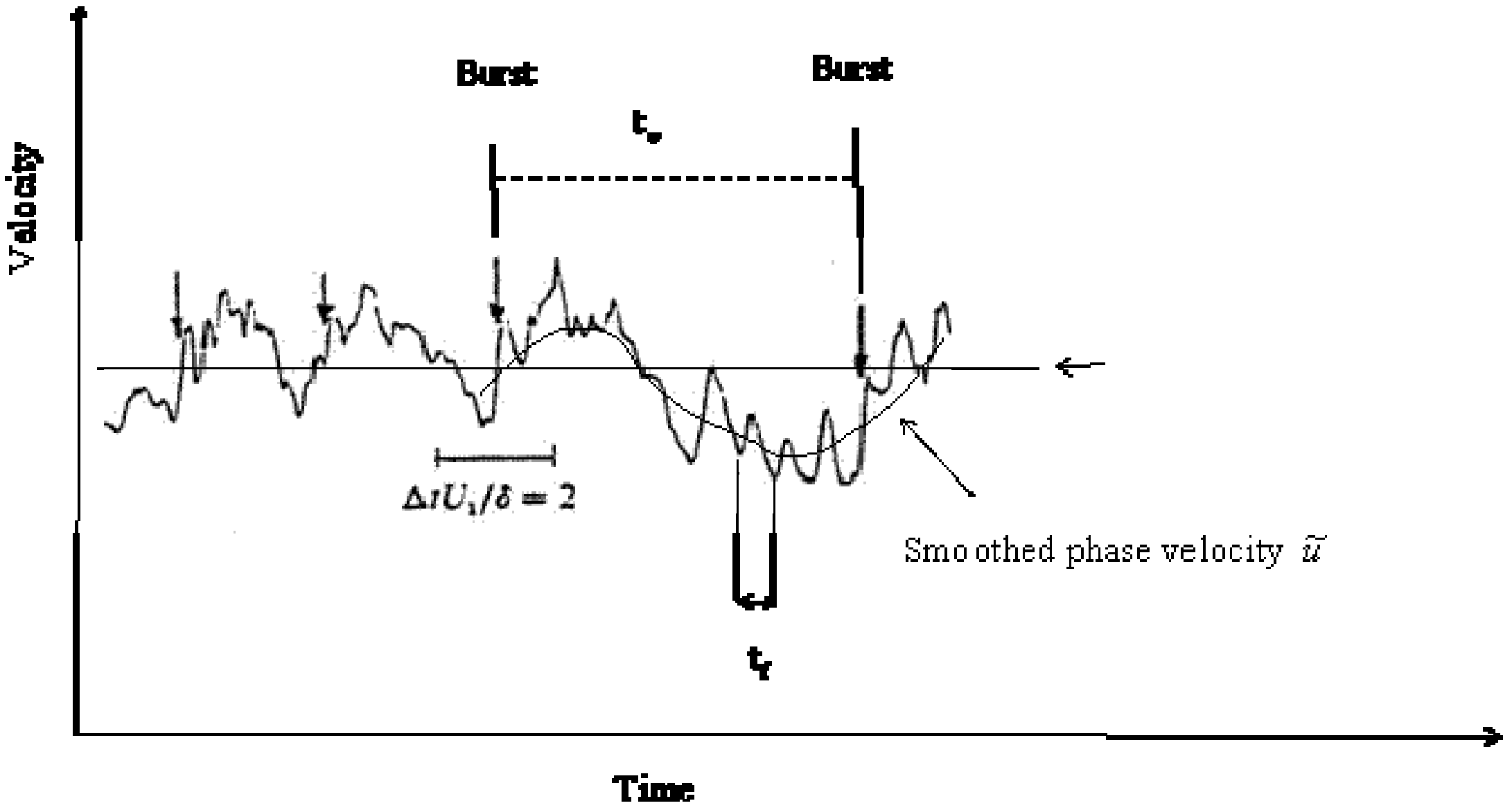

Figure 3 Decomposition of the instantaneous velocity in the sweep phase. Data of Antonia et al.(1990).

If we average over a longer period we obtain the Reynolds Averaged Navier Stokes equations RANS

$$U_i \frac{\partial U_j}{\partial x_j} = \nu \frac{\partial^2 U_i}{\partial x_i} - \frac{\partial \overline{U'_i U'_j}}{\partial x_i} \tag{8}$$

Where the instantaneous velocity $u_i$ at any point is decomposed into a long-time average value $U_i$ and a fluctuating term $U'_i$.

Thus

$$U'_i = \tilde{U}'_i + u'_i \tag{9}$$

$$\tilde{U}'_i = \tilde{u}_i - U_i \tag{10}$$

And then

$$u_i = U_i + \tilde{U}'_i + u'_i \tag{11}$$

Equation (7) defines a second set of Reynolds stresses $\overline{u'_i u'_j}$ which we will call "fast" Reynolds stresses to differentiate them from the standard Reynolds stresses $\overline{U'_i U'_j}$.

We may write the fast fluctuations in the form

$$u'_i = u_{0,i}\left(e^{i\omega t} + e^{-i\omega t}\right) \tag{12}$$

The fast Reynolds stresses $u'_i u'_j$ become

$$u'_i u'_j = u_{0,i} u_{0,j} (e^{2i\omega t} + e^{-2i\omega t}) + 2 u_{0,i} u_{0,j} \tag{13}$$

Equation (13) shows that the fluctuating periodic motion $u'_i$ generates two components of the "fast" Reynolds stresses: one is oscillating and cancels out upon long-time-averaging, the other, $u_{0,i} u_{0,j}$ is persistent in the sense that it does not depend on the period $t_f$. The term $u_{0,i} u_{0,j}$ indicates the startling possibility that a purely oscillating motion can generate a steady motion which is not aligned in the direction of the oscillations. The qualification steady must be understood as independent of the frequency $\omega$ of the fast fluctuations. If the flow is averaged over a longer time than the period $t_\nu$ of the bursting process, the term $u_{0,i} u_{0,j}$ must be understood as transient but non-oscillating. This term indicates the presence of transient shear layers embedded in turbulent flow fields and not aligned in the stream wise direction similar to those associated with the streaming flow in oscillating laminar boundary layers e.g. (Schneck & Walburn, 1976). Schoppa and Hussain (2002) have analysed their DNS data base of the wall layer and similarly argue that sinusoidal velocity fluctuations led to the production of intense shear layers associated with the streaming flow, that they call transient stress growth TSG. These terms are more easily visualised if we adopt the terminology used in traditional analyses of laminar oscillating flows e.g. (Schlichting, 1960; Stuart, 1966; Tetlionis, 1981). We define a stream function $\psi$ such that

$$u = \frac{\partial \psi}{\partial y} \qquad v = \frac{\partial \psi}{\partial x} \tag{14}$$

The basic variables are made non-dimensional

$$x^* = \frac{x}{L} \qquad y^* = \frac{y}{\sqrt{2\nu/\omega}} \qquad t^* = t\omega \tag{15}$$

$$U_e^*(x,t) = \frac{U_e}{U_\infty}(x,t) \qquad \psi^* = \psi\left(U_\infty\sqrt{\frac{2\nu}{\omega}}\right)^{-1} \tag{16}$$

where $U_\infty$ is the approach velocity for $x \to \infty$, $U_e$ is the local mainstream velocity and L is a characteristic dimension of the body. The NS equation may be transformed as:

$$\frac{\partial^2 \psi^*}{\partial y^* \partial t^*} - \frac{1}{2}\frac{\partial^3 \psi^*}{\partial y^{*3}} - \frac{\partial U_e^*}{\partial t^*} = \frac{U_e}{L\omega}\left(-\frac{\partial \psi^*}{\partial y^*}\frac{\partial^2 \psi^*}{\partial y^* \partial x^*} + \frac{\partial \psi^*}{\partial x^*}\frac{\partial^2 \psi^*}{\partial y^{*2}} + U_e^*\frac{\partial U_e^*}{\partial x^*}\right) \tag{17}$$

For large frequencies, the RHS of equation (17) can be neglected since

$$\varepsilon = \frac{U_e}{L\omega} << 1 \tag{18}$$

Thus to a first approximation, the stream function can be written as

$$\psi = \psi_0 \tag{19}$$

Equation (19) is accurate only to an error of order $\varepsilon$. Tetlionis reports a more accurate solution for the case when $\varepsilon$ cannot be neglected (i.e. for lower frequencies):

$$\psi = \psi_0 + \psi_1 + \psi_{st} + 0(\varepsilon^2) \tag{20}$$

The instantaneous velocity must then be written as

$$u_i = \tilde{u}_i + u'_i + u_{st} \tag{21}$$

This analysis and the evidence presented by Kline et al. and many DNS indicate that the wall layer, the log-law region and the far field region are based on completely different subsets of the NS equations reflecting different flow mechanism. It appears unreasonable then to expect that all three sub-regions would scale with the wall parameters ($u_*$, $\nu$) especially since the log law and the far field regions have no direct contact with the wall. Better parameters can be found *at the interfaces between these regions since they would be common to both adjacent regions*. Thus for the inner region, the thickness of the wall region $\delta_\nu$ and the velocity at its boundary $U_\nu$ looked much more attractive candidates.

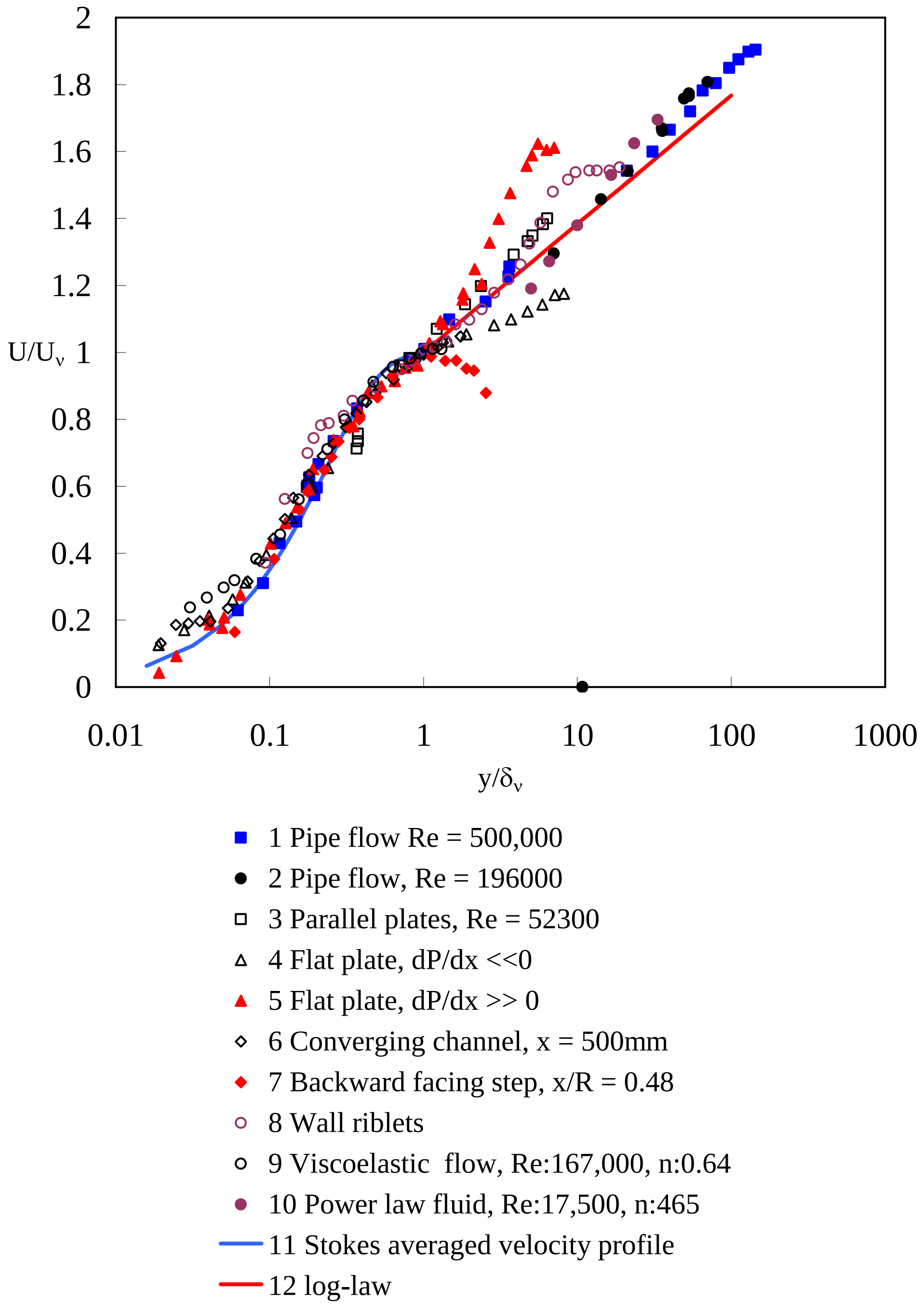

Figure 4 Zonal similar velocity profile for different types of fluids and flow configurations from Trinh (2005). Data from [1] Laufer (1954), [2, 10] Bogue(1962), [3] Schlinder & Sage (1953), [4,5] Kline, et al..(1967),[6] Tanaka & Yabuki (1986), [7] Devenport & Sutton (1991), [8] Bandyopadhyay (1986), [9] Pinho & Whitelaw (1990).

This approach, called zonal similarity analysis  produced a unique master curve for a large range of Reynolds numbers, different flow configurations and fluid rheological behaviours is shown in Figure 4 (K.T. Trinh, 1994, 2005; K. T. Trinh, 2009).

| Type of fluid | n | Re | $\delta_v^+$ | Source | Remarks |
|---|---|---|---|---|---|
| Newtonian | 1 | 2600 | 83.2 | Bogue (1962) | |
| | | 3300 | 75.0 | | |
| Viscous, non-Newtonian | 0.745 | 3660 | 80.1 | | Metzner-Reed (1955) generalised Reynolds number |
| | 0.70 | 11700 | 80.05 | | |
| | 0.59 | 6100 | 85.0 | | |
| | 0.53 | 17400 | 79.85 | | |
| | 0.465 | 7880 | 80.0 | | |
| Viscoelastic | 1 | 16700 | 60.0 | Pinho & Whitelaw (1990) | Reynolds number based on the non-Newtonian viscosity at the average wall shear stress |
| | 0.90 | 16700 | 105.0 | | |
| | 0.75 | 16700 | 155.0 | | |
| | 0.64 | 16700 | 180.0 | | |
| | | 459000 | 60 | Wells (1965) | Reynolds number based on the solvent viscosity |
| | | 98000 | 74 | | |
| | | 38700 | 82 | | |
| | | 211000 | 60 | | |
| | | 69900 | 75.6 | | |
| | | 13300 | 88 | | |

Table 1 Dimensionless wall layer thickness in pipe flow for different Reynolds

numbers and fluid properties

Table 1 shows how the thickness $\delta_\nu^+$ of the wall layer under different situations. The variation of this layer can be quite complex, for example in the recirculation region behind a backward facing step investigated by Devenport and Sutton(1991) . In particular its value is greatly increased by the occurrence of DR.

With increased polymer concentration, the thickness of the wall layer increases and the outer portion of the of the wall layer is now situated at distances traditionally attributed to the log-law region for Newtonian fluids. In addition, the portion of the profile between $0.1 < y^+ / \delta_\nu^+ < 0.8$ can be approximated by a straight line as shown in Figure 4.

$$\frac{U^+}{U_\nu^+} \approx 0.36 \ln\left(\frac{y^+}{\delta_\nu^+}\right) + 1.1 \qquad (22)$$

At the end of the laminar region for Newtonian flow, $\mathrm{Re} = 2100$ $R^+ = \delta_\nu^+ = 64.7, \quad U_\nu^+ = R^+/2$ and equation (22) becomes

$$U^+ = 11.7 \ln y^+ - 18 \qquad (23)$$

which almost coincides perfectly with Virk's asymptote.

If we match equation (22) with the edge of the wall layer at high Reynolds numbers in Newtonian flow $\delta_\nu^+ = 64.7, \quad U_\nu^+ = 15.6$ (K. T. Trinh, 2009) we obtain

$$U^+ = 5.6 \ln y^+ - 6.24 \qquad (24)$$

which is very similar to Karman's correlation for the buffer layer.

The velocity in the sweep phase has been modelled with Stokes'solution for a suddenly started flat plate (Einstein & Li, 1956; Hanratty, 1956; Meek & Baer, 1970). However, the governing equation should not be expressed in terms of the smoothed phase velocity $\tilde{u}_i$, not the instantaneous velocity $u_i$ used by Einstein and Li:

$$\frac{\partial \tilde{u}}{\partial t} = \nu \frac{\partial^2 \tilde{u}}{\partial y^2} \qquad (25)$$

Stokes (1851) gives the solution as

$$\phi = erf(\eta_s) \qquad (26)$$

where $\eta_s = \frac{y}{\sqrt{4\nu t}}$, $\phi = \frac{\tilde{u}}{U_\nu}$

Equation (26) describes the penetration of wall viscous momentum into the main flow and $\delta_\nu$ represents the maximum distance of penetration. It applies only to the sweep phase because the streaming flow reaches much further into the inviscid flow (K. T. Trinh, 2009). However since the sweep phase lasts so much longer than the bursting phase it dominates the average velocity distribution in the wall layer (Walker, Abbott, Scharnhorst, & Weigand, 1989). The error function in equation (26) is averaged over low speed streaks of all possible ages that pass the measuring probe. The resulting curve, shown in Figure 4, fits the experimental data quite adequately. Thus in this visualisation, the Virk's asymptote cannot be interpreted as a new log-law but as the outer portion of the wall layer. There is no argument here about the role of dissolved polymers: they do make the wall layer much thicker. the elasticity of the flow medium in viscoelastic fluids dampens the magnitude of the oscillations (fast velocity fluctuations) $\omega$ and slows down their rate of growth. As a consequence the magnitude of the fast Reynolds stresses (strength of the streaming flow) is reduced, the ejection of the streaming flow is delayed and the sweep phase lasts longer. In fact any technology that dampens the fast oscillations such as riblets, bubbles or compliant skins similarly extends the thickness of the wall layer and produces DR (K. T. Trinh, 2009).

By its very nature, the error function erf can be approximated by a straight line for small values of $\eta_\nu$ i.e. very close to the wall. Such linear relationship was observed experimentally by Popovich and Hummel (1967). We can approximate the time averaged equation (26) with $\overline{\phi} = \overline{\eta}_s / 4.14$ up to a value $\overline{\eta}_s = 0.07$. In Newtonian fluids, with $\delta_\nu^+ = 64.7$, $U_\nu^+ = 15.6$ this leads to a limit of $y^+ = U^+ = 4.5$. (K. T. Trinh, 2009). This does not mean that the flow in the so-called laminar sub-layer is purely viscous in the steady state sense envisaged by Prandtl and Virk (op.cit.). The point $U^+ = y^+ = 11.6$ simply represents the intersection of equations (2) and (3); it is not physically real. Equation (2) does not fit any known Newtonian velocity profile in the range $4.5 < y^+ < 11.6$. For the wall thickness of $\delta_\nu^+ = 180$ in the data of Pinho at DR=64% (Table 1) this limit becomes $y^+ = U^+ = 12.6$. Thus the postulate of a new elastic sub-layer starting at the point (11.6,11.6) is not warranted. Virk's asymptote should instead be compared with Karman's profile of the buffer layer(Karman, 1934). Karman estimated this layer to extend between $5 < y^+ < 30$ for Newtonian fluids. This limit of the Karman buffer layer can be shown to be the time-averaged value of maximum penetration of viscous momentum into the main flow (K. T. Trinh, 2009). In DR flows since the wall layer (the maximum distance of penetration of viscous wall momentum) is much thicker, the buffer layer is similarly extend. For the data of Pinho quoted here, the buffer layer is estimated to extend to a distance $y_b^+ = 83.5$ which agrees well with the intersection between the data of Pinho and Virk's asymptote plotted in Figure 1.

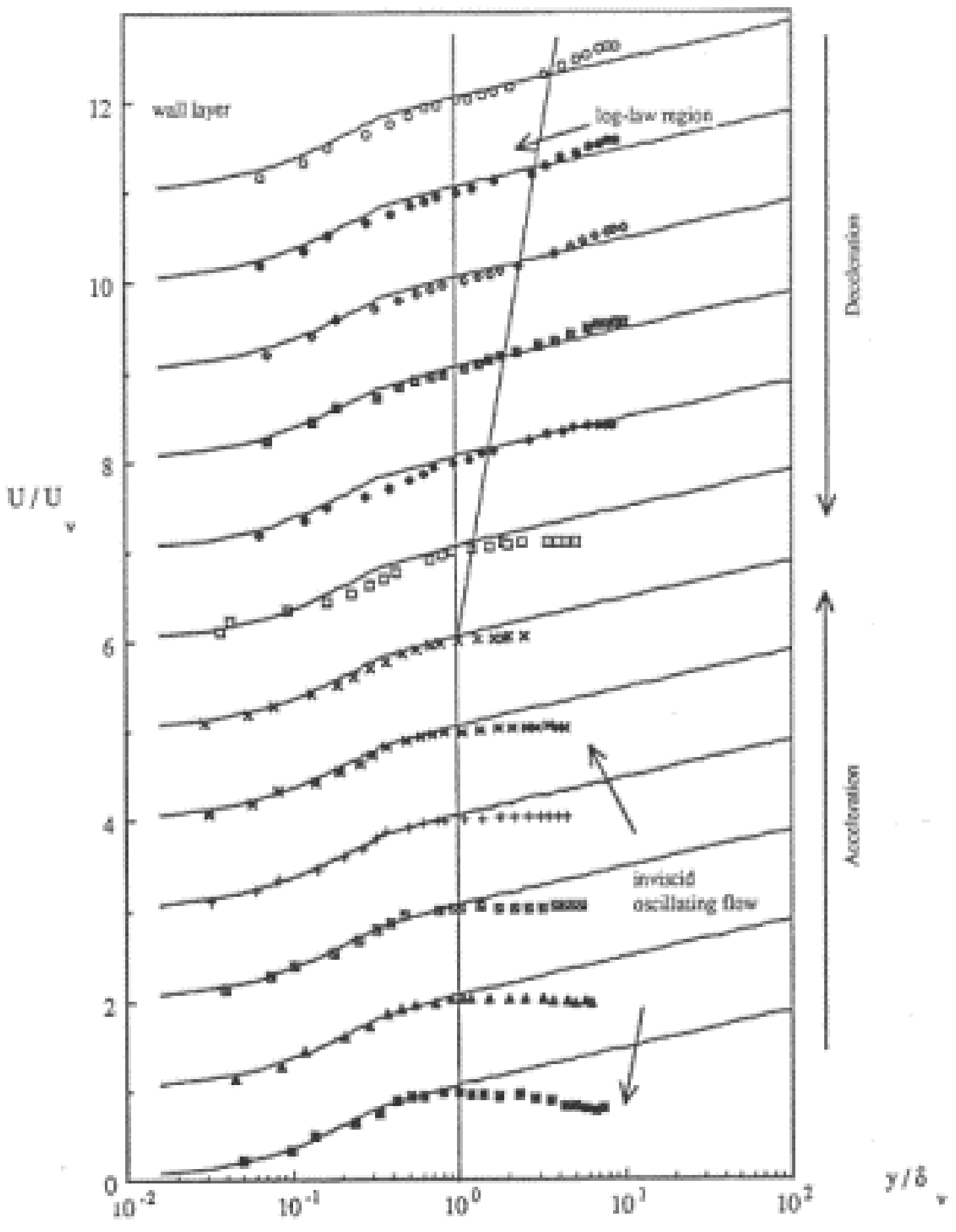

Figure 5 Penetration of the log-law into the outer region during a cycle oscillating pipe flow. From Trinh (2009). Data of Akhavan et al.. (1991) representing velocity profiles at different times in a cycle of oscillating flow. The log-law in the turbulent flow field only begins to grow in the deceleration phase.

The error function is the simplest and most convenient mathematical model for the wall layer profile but it is not the best (Trinh 2009). In particular it assumes that the velocity gradient is zero at the wall layer edge, which is not true. Thus it is not a reliable model for $\overline{\eta}_s > .95$.

A final point worth discussion is Virk' maximum DR friction factor curve which is clearly different from Poiseuille's equation for laminar pipe flow (Figure 2). The ejections associated with bursting have been compared to transient jets of fluid essentially in cross flow to the main stream (Grass, 1971; Townsend, 1956). The fluid in the jet moves as coherent structure as evidenced in the analysis of both DNS data (Johansson, Alfredson, & Eckelmann, 1987) and probe measurements (Brown & Thomas, 1977). The jet in crossflow has been divided into three zones: the linear near-field region is jet dominated in the sense that the effects of the crossflow on the jet are not yet significant; in the curvilinear region, the initial jet momentum and the momentum extracted from the crossflow have comparable effects on the jet characteristics; in the far-field region, the effects of the crossflow predominate and the jet is aligned in the direction of the crossflow. The far field region corresponds to the outer region of turbulent flow fields and has first been modelled by Coles (1956) by a law of the wake. The linear region corresponds to the log-law which, as Millikan (1939) showed is a necessary transition between a wall layer scaled with the wall variables and the outer region scaled with the outer variables.

It is shown elsewhere (K. T. Trinh, 2009) that at the transition between laminar and turbulent flow (Re=2100) the jets are still weak and only the far field region protrudes beyond the wall layer. Thus at transition the log-law does not yet exist and the flow field consists only of a wall layer and an outer region. The log-law sub region grows as the Reynolds number increases and turbulence becomes stronger. This growth is well illustrated by replotting the data of Akhavan et al. (1991) who studied the transition between laminar and turbulent flow within one cycle of oscillatory pipe flow.

Virk is absolutely correct in his observation that DR decreases the extent of the log-law sub-region. The presence of polymers dampen the fast fluctuations thus decreasing the strength of the streaming flow and hence its penetration beyond the wall layer. This phenomenon occurs in all DR flows, not just polymer solutions. The maximum DR asymptote in Figure 2 is linked with velocity profiles made up of a wall layer plus a law of the wake region. This can be seen clearly in the data of Pinho in Figure 1. It does not mean however that the fast fluctuations and the streaming flow cannot be dampened further to the point of suppressing the law of the wake sub-region too. Then the flow becomes laminar. Thus, in my views, the maximum Dr achievable by polymer addition is not Virk's asymptote but a full relaminarisation of the flow field.

## Conclusion

Virk's asymptote does not represent a new log-law with a modified mixing-length. This log-law is similar in nature to Karman's buffer layer profile and reduces to it for Newtonian fluids. Thus it is simply part of the wall layer velocity profile but is extended because of the increase in wall layer thickness in DR flows. The friction factors at the maximum drag reduction asymptote correspond to the velocity profiles consisting of a wall layer and a law of the wake sub-region. Maximum drag reduction results in a the suppression of the law of the wake and full relaminarisation of the flow.